\documentclass{article}

\pdfoutput=1

\usepackage{natbib}
\usepackage{graphicx}
\usepackage{times}

\usepackage{url}             

\newcommand{\degr}{\ensuremath{^\circ}}

\newcommand{\rsun}{\ensuremath{R_\odot}}

\newcommand{\aap}{    {\it Astron. Astrophys.}}

\newcommand{\apj}{    {\it Astrophys. J.}}

\newcommand{\grl}{    {\it Geophys. Res. Lett.}}

\newcommand{\mnras}{  {\it Mon. Not. Roy. Astron. Soc.}}
\newcommand{\nat}{    {\it Nature}}

\newcommand{\solphys}{{\it Solar Phys.}}

\newcommand{\ssr}{    {\it Space Sci. Rev.}}

\begin{document}

    \title{The Extended Solar Cycle Tracked High into the Corona.}
    \author{S.J. Tappin\footnote{National Solar Observatory, Sacramento Peak, Sunspot,
      NM, USA, email: \protect\url{jtappin@nso.edu}}, R.C. Altrock\footnote{Air Force Research Laboratory, Space Weather Center of Excellence,
      National Solar Observatory, Sacramento Peak, Sunspot, NM, USA, email:
      \protect\url{altrock@nso.edu}}}
\maketitle{}
\markboth{Tappin \& Altrock}{Extended Solar Cycle in the High Corona.}

    \begin{abstract}
      We present observations of the extended solar cycle activity in
      white-light coronagraphs, and compare them with the more familiar
      features seen in the Fe~\textsc{xiv} green-line corona. We show that the
      coronal activity zones seen in the emission corona can be tracked
      high into the corona. The peak latitude of the activity, which
      occurs near solar maximum, is found to be very similar at all
      heights. But we find that the equatorward drift of the activity
      zones is faster at greater heights, and that during the declining
      phase of the solar cycle, the lower branch of activity (that
      associated with the current cycle) disappears at about 3\rsun{}.
      This implies that that during the declining phase of the cycle,
      the solar wind detected near Earth is likely to be dominated by
      the next cycle. The so-called ``rush to the poles'' is also seen
      in the higher corona. In the higher corona it is found to start
      at a similar time but at lower latitudes than in the green-line
      corona. The structure is found to be similar to that of the
      equatorward drift.
    \end{abstract}

\section{Introduction}
\label{sec:introduction}

The existence of an ``extended solar cycle'' in which features related
to the next cycle appear at high latitudes around or soon after the
maximum of the current cycle has been known on the solar surface and in
the low corona for many years (\textit{e.g.}\ \citealp{labonte82};
\citealp{wilson88}; \citealp{altrock97}).  Typically this activity is
observed in the form of ephemeral active regions, torsional waves and
coronal streamers observed in the Fe~\textsc{xiv} green line. Having appeared
near solar maximum, these structures then drift equatorwards mirroring
the ``butterfly'' pattern of sunspots and active regions at lower
latitudes in the current cycle. Following solar minimum, the sunspots
and active regions reappear at the same latitudes as this descending
coronal and rotational structure. At this time a zone of coronal
activity detaches from the main zone and moves rapidly poleward,
joining the next extended cycle around solar maximum, this is
described as the ``rush to the poles'' \citep{waldemeier64},
(see also \citet{altrock11} for examples).

To date, studies of the extended cycle and of the rush to the poles
have been confined to the solar surface and the low corona.  But now
that observations from the LASCO coronagraphs on SOHO
\citep{brueckner95} are available for more than a complete solar cycle
it is possible to carry out an investigation of how these structures
propagate into the higher corona.

In this paper we combine the Fe~\textsc{xiv} green-line and white-light
coronagraph observations to trace the evolution of the zones of
activity in latitude as a function of height above the solar limb and
phase of the solar cycle. 

\section{The Observations}
\label{sec:observations}

In this paper we make use of five coronagraph datasets to track
activity in the corona from 1.15\rsun{} to 20\rsun{} (N.B.\ in this
paper, all heights are measured from Sun centre).

\subsection{NSO Green Line}
\label{sec:obs-nso-green}

The Fe~\textsc{xiv} 530.3~nm green line coronagraph observations made at the
National Solar Observatory at Sacramento Peak (NSO/SP), have been
discussed in some detail by \citet{altrock97}. In this study we
have used those data from 1986 to the end of 2011. In summary: On each day with
sufficiently good observing conditions a scan of the corona is made at
a distance of 1.15\rsun{} from disk centre with an angular resolution
of 3\degr{} in position angle, and a scan width of 1.1~arcmin.

\subsection{MLSO Polarised White Light (Mk3 and Mk4 Coronameters)}
\label{sec:obs-mlso-pb}

The Mauna Loa Solar Observatory (MLSO) has been making white-light
polarised brightness images since 1980. The Mk3 coronameter
\citep{fisher81} was operated from 1980 to 1999, and the Mk4
\citep{elmore03} from 1998 to the present. For both instruments a linear
detector array is scanned around the corona to build up an image in
position angle and radial distance. For the Mk3 instrument a few scans
were made each day (during the interval considered in this paper there
are typically two or three images on the MLSO data download site) while for
the Mk4 scans are taken every three minutes through the 5-h
observing day. These scans are published at an angular resolution of
0.5\degr{} in position angle by 10~arcsec in the radial direction for
Mk3, and 5~arcsec for Mk4. Both instruments cover similar regions of
the corona, 1.12\rsun{} to 2.44\rsun{} for Mk3 and 1.12\rsun{} to
2.86\rsun{} for Mk4. For this study we have used the daily average
images for all available days from 1986 to 2011 (for 2009 and 2010,
where many daily average images are missing, we have downloaded the
full dataset and generated daily average images from the individual
scans). During the interval of overlap we use the Mk4 image if both are
available. At the time of writing the most recent data available were
for 19~September~2011.

\subsection{LASCO White Light (C2 and C3)}
\label{sec:obs-lasco}

The LASCO coronagraphs on SOHO \citep{brueckner95}, have been making
regular images of the corona since late 1995. For this study we use the
C2 coronagraph which covers the range from about 2\rsun{} to 6\rsun{},
and the C3 coronagraph which covers the range from about 4\rsun{} to
30\rsun{}. For C2 the images are taken using an orange filter with a
cadence of about 20~min for most of the mission. C3 has a cadence
of around 30~min and uses a clear filter for its primary synoptic
sequence.At the time of writing the most recent data available were
for 1~April~2011.

\section{Data Processing}
\label{sec:data-proc}

The first step in the data processing was to reduce the data to a
common coordinate system and resolution. For this we adopted the
resolution of the NSO Fe \textsc{xiv} dataset, which has the lowest resolution
in position angle of the five instruments.

For the MLSO data which were already in polar coordinates, all that
needed to be done was to average together position angle bins to reduce
the resolution from 0.5\degr{} to 3\degr{}, and average radially over
an annulus approximately 70arcsec wide ($\pm 35$arcsec from the
nominal radius).

Unlike MLSO where  daily (\textit{i.e.}\ 5~h) average images are available,
LASCO data are only available as individual frames. Therefore to
improve the signal to noise and also to make any time-smearing effects
comparable to those in the MLSO data, we first read a full day of
images and made exposure corrections by the method described by
\citet{tappin99}. We then generated four 6-hour average images for
each day.

Since LASCO uses a camera with a rectangular CCD rather than a rotating
linear detector, the raw images are in cartesian coordinates. We
therefore needed to convert the image to polar coordinates. To do this,
the field of view was divided into radial bins of width equal to four
pixels in the original data (47.6~arcsec for C2 and 224~arcsec for C3)
and into 3\degr{} sectors in position angle. For each resulting cell,
the pixels whose centres lie in the cell were determined and the count
rates in those pixels were averaged.

\begin{table}
  \begin{tabular}{ll}
    \hline
    Observatory & Heights \\
    \hline
    NSO/SP & \textbf{1.15} \\
    MLSO (both) & \textbf{1.15}, 1.20, 1.25, 1.30, \textbf{1.4}, 1.5,
    1.6, \textbf{1.7}, 1.8, 1.9, 
    \textbf{2.0}, 2.2, \textbf{2.35} \\
    LASCO (C2) & \textbf{2.5}, 2.75, \textbf{3.0}, 3.25, \textbf{3.5},
    3.75, \textbf{4.0} \\ 
    LASCO (C3) & 5.0, \textbf{6.0}, 7.0, 8.0, 9.0, \textbf{10.0}, 12.0,
    \textbf{14.0}, 16.0, 18.0, \textbf{20.0} \\ 
    \hline
  \end{tabular}
  \caption{The heights of the scans used in this study. All distances
    are in units of \rsun{} from the centre of the Sun. The heights
    which are shown in Figure~\ref{fig:activity96} are indicated by
    \textbf{bold} font.} 
  \label{tab:scans-list}
\end{table}

We then selected a number of radii at which to extract circumferential
profiles covering the range from 1.15 to 2.35\rsun{} for the MLSO data,
2.5 to 4.0\rsun{} for LASCO C2 and 5 to 20\rsun{} for LASCO C3. The
full list of heights used is shown in Table~\ref{tab:scans-list}.  It
should be noted that prior to July 1997, the majority of LASCO images
were not full-field. As a result the C2 scans above about 3.5\rsun{}
and C3 scans above about 18\rsun{} have a lower signal-to-noise
over the poles prior to that date.

Because the different position angles in the LASCO images were recorded
with different parts of the detector, we found that variations in the
detector dark current resulted in certain latitudes at which persistent
maxima and minima occurred. Since these instrumental features make it
difficult to see the maxima due to solar wind structures, it was
necessary to find a way to remove them. We have found that an effective
method is to divide the data into one year blocks so as to have a long
enough base, but also to handle the variations in the CCD properties
with time.  For each year, we determine the fifth percentile of data at
each location in the scans, and subtract the resulting value from all
the scans for that year. This is chosen in preference to using an
annual minimum as there are occasional corrupt images which are likely
to dominate the minimum.

In 2004, the Z-axis drive of SOHO's high-gain antenna failed. Since
that time the spacecraft has performed a 180\degr{} roll manoeuvre four
times a year to keep Earth in the antenna beam. This was corrected
after the CCD pattern removal to align the scans correctly.

For all the datasets we then followed \citet{altrock03} and
defined a significant maximum as one in which the brightness at a
position angle was a local maximum, and also its immediate neighbours
exceeded the next pair of position angles (\textit{i.e.}\ for position angle
index $i$ to be considered a significant maximum then
\begin{equation}
  b_{i-2} < b_{i-1} < b_i > b_{i+1} > b_{i+2}
\label{eq:max_def}
\end{equation}
must be satisfied; $b_i$ denotes the observed intensity at position
angle index $i$). As well as excluding noise, this criterion also
eliminates most maxima due to stars and planets (even Venus in LASCO C3
does not extend over 15\degr{}). This method does not distinguish
bright and faint maxima, but rather gives us a measure of where
streamers are present. To adjust the observation dates of the two limbs
to central meridian, maxima over the East limb were assigned a date seven
days later than the date of observation, while those over the West limb
were given a date seven days earlier. We then convert the position angles
of the maxima to latitude (on the assumption that the features lie
above the limb). For LASCO we excluded the quadrant containing the
occulter support pylon as this region is vignetted by the pylon thus
compromising our ability to detect maxima. The maxima for all four (or three
for LASCO) quadrants were then combined to give a list of maxima
against absolute latitude. We consider that the gain in signal to noise
thus obtained outweighs any loss of information about North-South
asymmetries (\textit{c.f.}\ \citealp{altrock11}). All the quantitative fits
presented in this paper are made by fitting to this list of maxima.

For the purposes of displaying the data we then counted the number of
maxima that were detected in each latitude bin for each Bartels (27
day) rotation, and divided by the number of scans or images in the
rotation to give a probability of a maximum at that latitude.  To
facilitate visualisation we smooth the counts of maxima using a nine
rotation running mean, which was chosen as giving the best apparent
balance between signal to noise and smearing. 

\section{Results}
\label{sec:results}

\begin{figure}[tbp]
  \centering
  \includegraphics[width=\textwidth]{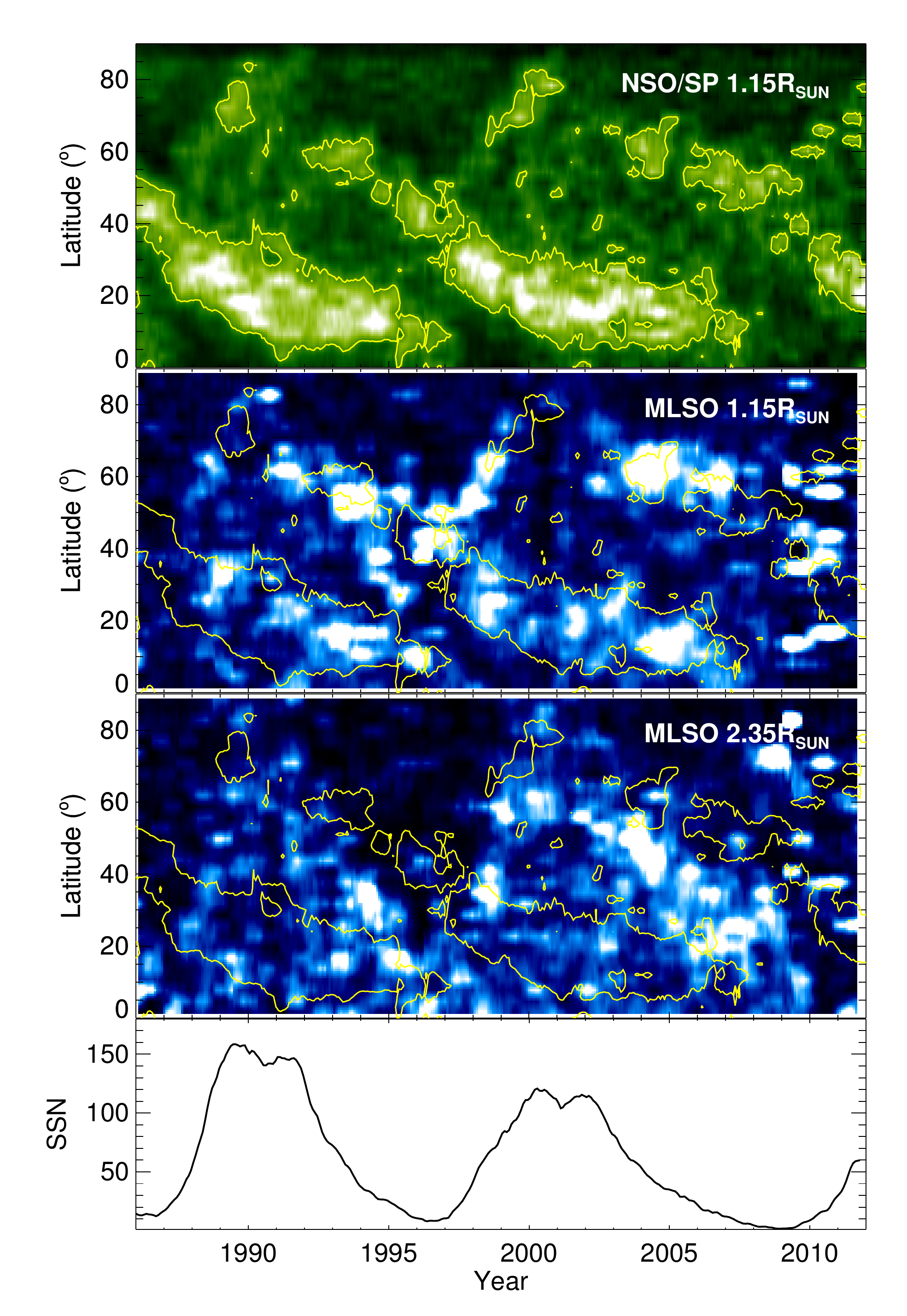}
  \caption{The evolution of the zones of activity from 1986 to
    2012. The panels show (from top to bottom) the NSO green line data
    at 1.15\rsun{}, the MLSO white-light data at 1.15\rsun{}, the MLSO
    data at 2.35\rsun{}, and the smoothed sunspot number. A single
    contour of the NSO green line map is shown in yellow over each
    panel to facilitate comparison. N.B.\ all heights are measured from
    the centre of the Sun.}
  \label{fig:evolution1986}
\end{figure}

In Figure~\ref{fig:evolution1986}, we show the NSO green line activity
map and two of those for the MLSO white light polarised brightness for
the interval 1986 to 2011 (solar cycles 22 and 23). These maps show the
number of maxima as a function of latitude and date. We remind
the reader that this does not show the amplitude of the maxima,
only the number. All three maps show a similar structure with two
overlapping cycles drifting towards the equator, and also the rush to
the poles.  The most important feature to notice here is that the NSO
map at 1.15\rsun{} and the MLSO map at the same height match very
closely although the latter is considerably noisier. Since the
K-coronal intensity is a function of density only, while the green line
emission is a function of both density and temperature, the close match
shows that the activity zones must be primarily density rather than
temperature features.
The MLSO map at 2.35\rsun{}, shows
a similar structure to the maps at 1.15\rsun{}, but as can be seen by
comparing it with the overlaid contour of the green line activity it is
very clear that the movement of the activity towards the equator during
the declining phase of both cycles is much faster than at 1.15\rsun{}.

\begin{figure}[tbp]
  \centering
  \includegraphics[width=\textwidth]{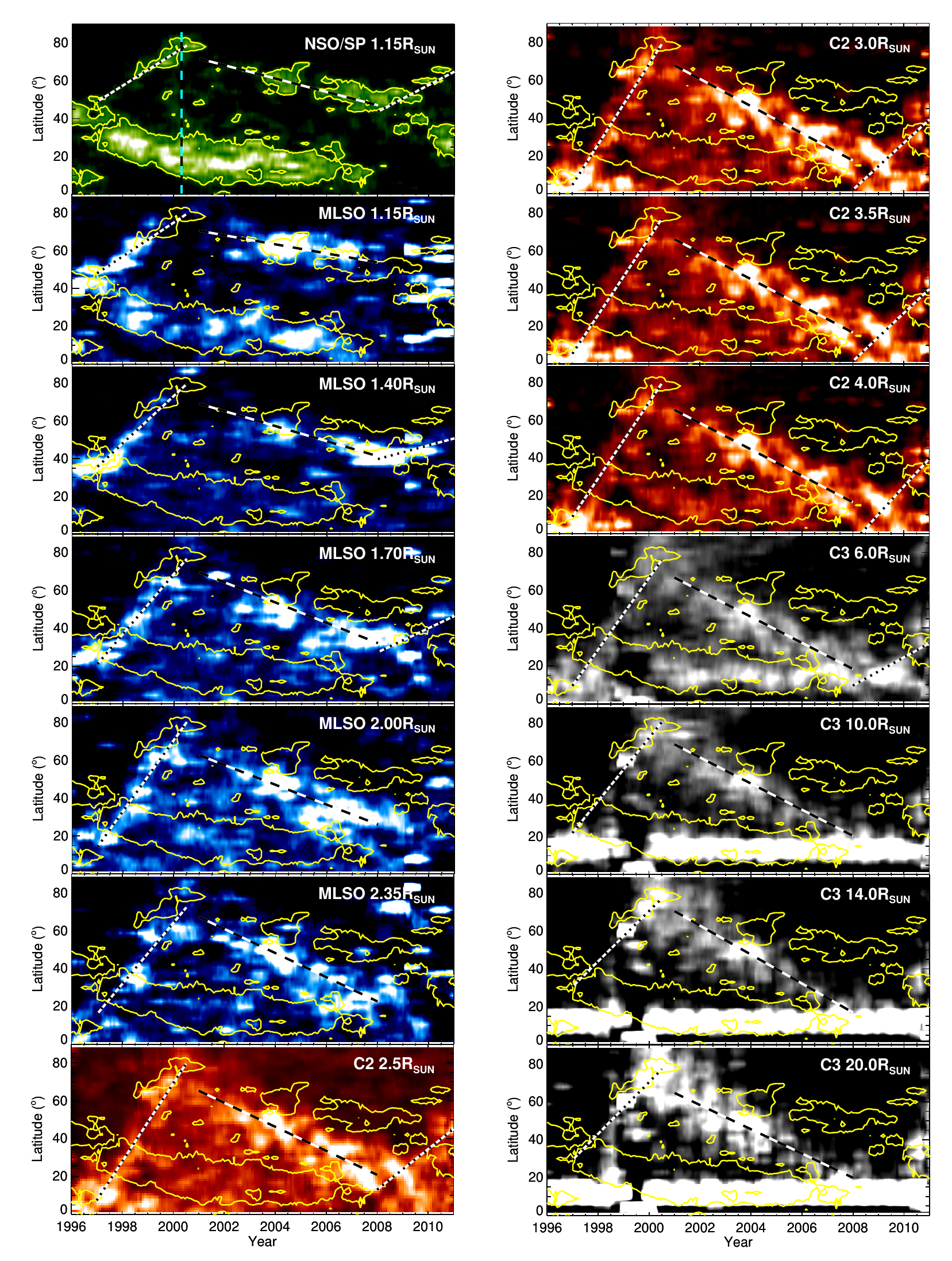}
  \caption{The evolution of the zones of activity for selected heights
    in the corona, from 1996 to 2011 (Fe\textsc{xiv} in green MLSO in
    blue, LASCO C2 in red, LASCO C3 in grey, respectively). For each
    height, a single contour from the NSO green line map (yellow) is
    superposed to facilitate comparison. Also superposed on each map is
    an estimate of the gradient of the drift towards the equator (black
    and white dashed lines), and of the ``rush to the poles'' (black
    and white dotted lines, black on white for cycle 23, white on black
    for cycle 24). The rush to the poles for cycle 24 is only shown for
    those heights at which it can be clearly discerned. The persistent
    maxima seen below 20\degr{} latitude in the LASCO C3 plots, the
    final four panels with heights of 6\rsun{} and above, is an effect
    of the F-corona which we have not been able to eliminate. The date
    of solar maximum is indicated by the cyan and black dashed line
    overlayed on the NSO map (upper left panel).}\label{fig:activity96}
\end{figure}

For a more quantitative analysis of the variation of the latitude of
the activity zones with height, we concentrate on the interval for
which we also have LASCO data, namely 1996-2010 which corresponds to
solar cycle 23. A selection of these datasets are shown in
Figure~\ref{fig:activity96}.

It should be noted that the constant band of ``activity'' below
20\degr{} latitude in LASCO C3 is due to the F-corona which dominates
at these altitudes. Even though the CCD irregularity removal does also
reduce this contribution, the small changes in apparent width and
inclination of the F-corona as the Earth orbits the Sun still produce a
dominant contribution to the number of detected maxima. Unfortunately
this means that we cannot track the activity bands below about
20\degr{} latitude in C3.

\begin{figure}
  \centering
  \includegraphics[width=\textwidth]{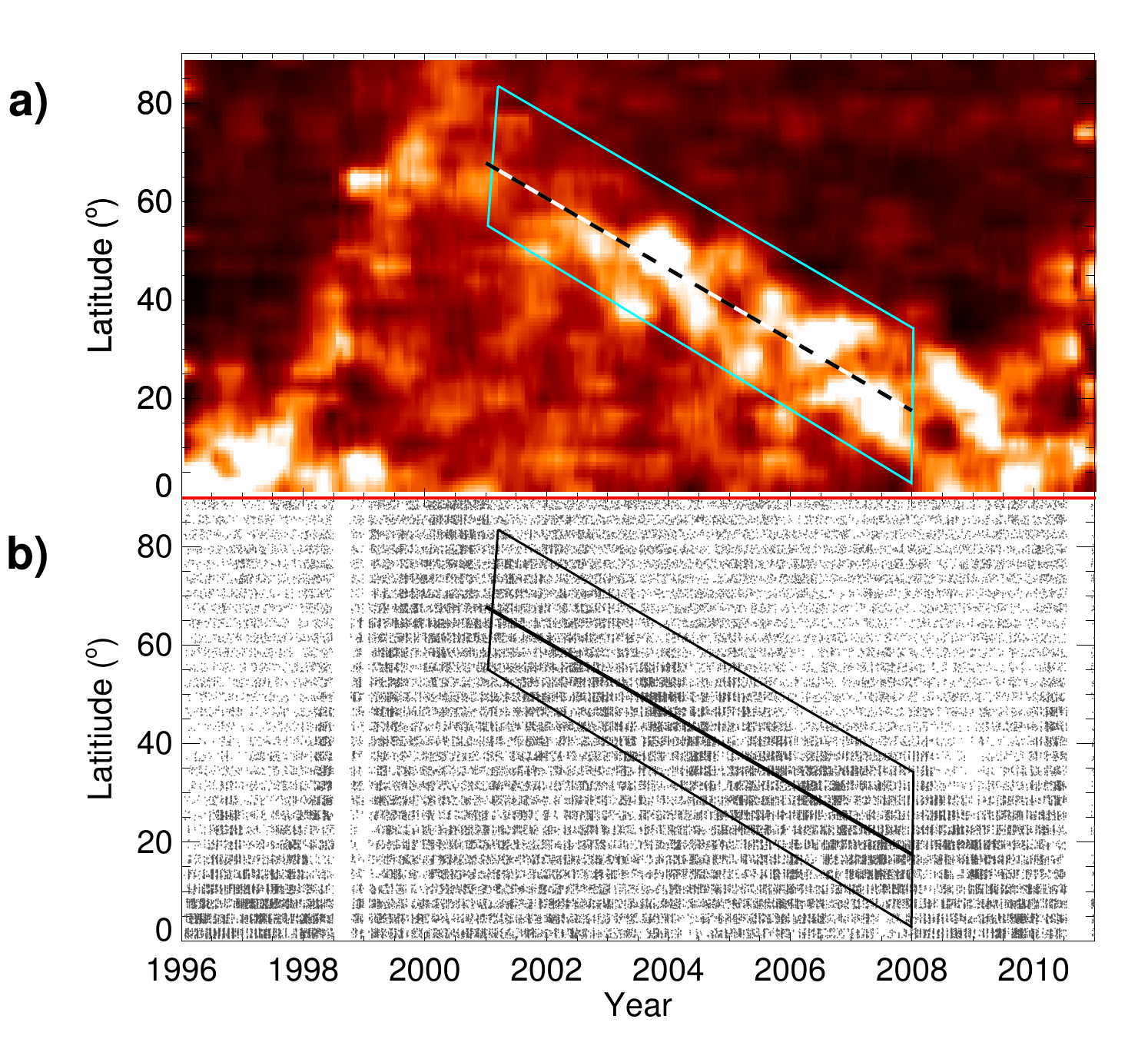}
  \caption{The process of fitting the trends to the drift. The example
    here is for LASCO C2 at 3.0\rsun. (a) The activity map and (b) The
    raw maxima (N.B. in the figure a random number between -1 and +1 is
    added to the latitude of each point to separate the individual
    points and allow the reader to see the variation of point density
    more easily). In each case the region of the equatorward drift is
    outlined with a medium weight box, and the fit is shown with a
    heavy line. The fit is generated by a linear regression to all the
    points within the region-of-interest box.}
  \label{fig:fitting}
\end{figure}
For the purposes of this study, we have considered only the interval
from 2001 to the start of 2008, where (a) the equatorward drift is
approximately linear in time and (b) it can be clearly seen in all the
instruments at all heights (we therefore do not consider the abrupt
acceleration of the drift from about the start of 2009, which is also
the time at which the cycle 24 rush to the poles separated from the
equatorward drift).  To quantify the trend in the equatorward
drift we firstly define manually a region which contains the activity
band of the drift, this is done simply by examining the activity maps
and drawing a box around the band of activity. We then perform a
linear fit to all the recorded maxima within that region to obtain a
slope and intercept. The chosen region and the comparison of
the smoothed map and the individual maxima is shown in
Figure~\ref{fig:fitting}. This method was chosen in preference to
fitting to the maxima at each time (as was done by
\citet{altrock03} for seven-rotation averages) since it is less
sensitive to spurious maxima.  The fits thus derived are indicated in
Figure~\ref{fig:activity96} as dashed black and white lines.

\begin{figure}[tbp]
  \centering
  \includegraphics[width=\textwidth]{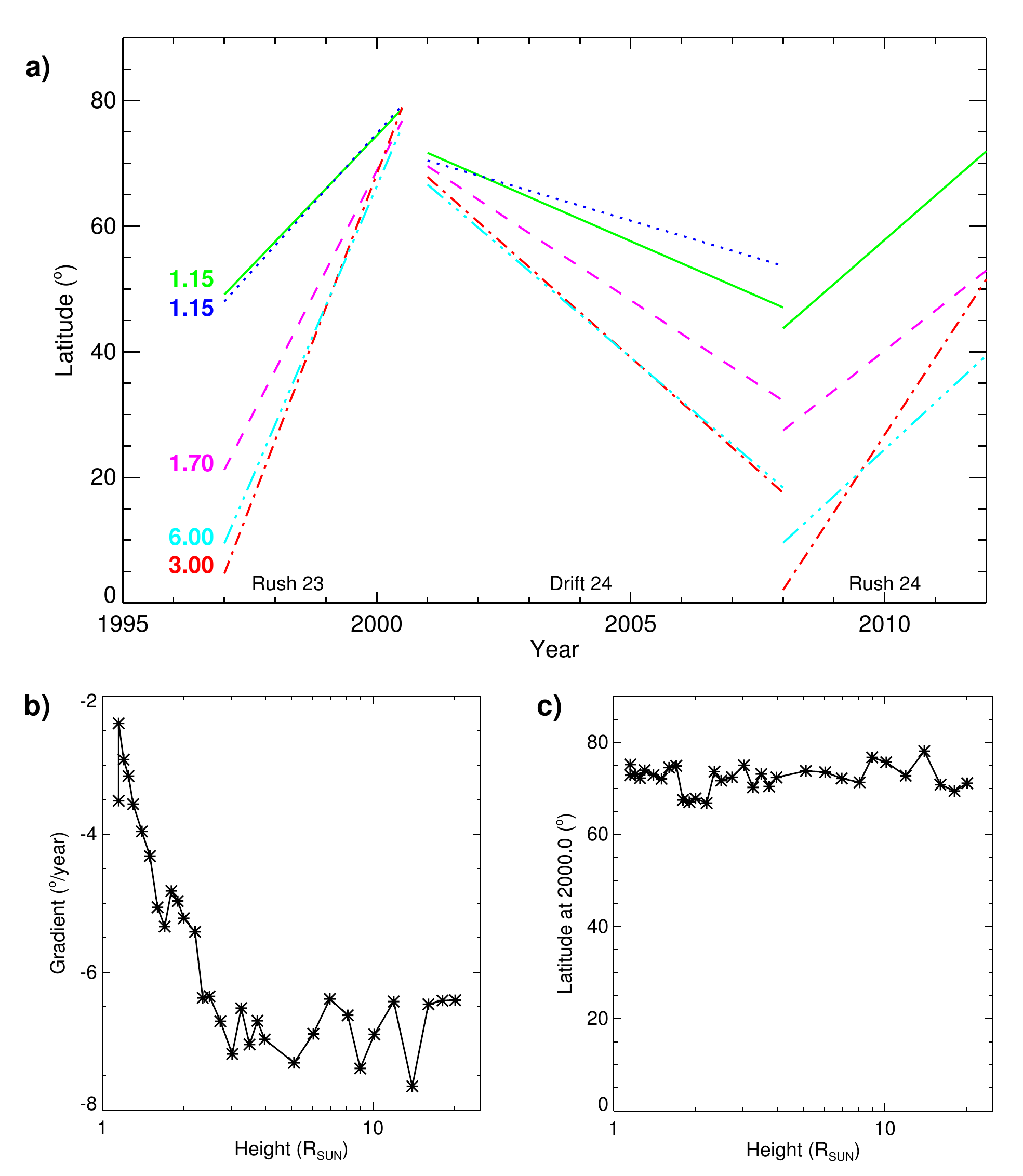}
  \caption{(a) The trends of selected equatorward drifts, and the
    rushes to the poles as overlayed on
    Figure~\ref{fig:activity96}. The heights are indicated to the left
    of the tracks for the cycle 23 rush to the poles, for the two
    trends at 1.15\rsun{}, the solid green lines are the NSO green-line
    data and the dotted blue lines are the MLSO data. Note that the
    rush of cycle 24 could not be discerned for the MLSO data at
    1.15\rsun{}. (b) The variation of equatorward drift rate with
    height as determined from the slope estimates in
    Figure~\ref{fig:activity96}. (c) The variation of latitude of the
    start of the equatorward drift at 2000.0 with height.}
  \label{fig:trends}
\end{figure}

It is clear from the activity zone maps in Figure~\ref{fig:activity96}
that at all heights, the coronal activity zones reach a maximum
latitude of around 75\degr{} near solar maximum in 2000 and then the
latitude of the activity zones moves toward the equator through the
declining phase of the cycle. However the drift to the equator is
clearly much faster at higher altitudes, and by about 4\rsun{} there is
little if any overlap between the cycles. To show this more clearly, we
overlay all the fitted trends from Figure~\ref{fig:activity96} in
Figure~\ref{fig:trends}a.

We plot the gradients of the estimates of the drift to the equator in
Figure~\ref{fig:trends}b, along with the latitude at the start of 2000
in Figure~\ref{fig:trends}c. This shows clearly that close to the Sun,
the drift rate is around 3\degr{}/year, but that this increases to
about 7\degr{}/year by 3-4\rsun{}, above that height the gradient is
constant within the errors of measurement. There is no clear trend in
the latitude at the start of 2000, with a value close to 75\degr{} at
all heights. We do not consider that the slight fluctuations are
significant.

We note that there are a number of epochs during which high activity is
seen at all heights. Most notable of these features is the enhanced
activity in late 2003 and early 2004. This combined with the smooth
change of latitude with height provides good evidence that we are
observing the same structures at all heights.

The other main component of the extended cycle, the so-called ``rush to
the poles'' which begins soon after solar minimum and reaches its
highest latitude near to solar maximum, is also clearly
visible at all heights. Historically the rush to the poles has been
tracked from about 45\degr{} latitude at solar minimum up to high
latitudes at maximum \citep{waldemeier64,altrock97}. However we see that
higher in the corona the rush starts at about the same time as it does
in the green-line measurements, but it begins at much lower
latitudes.  The trend that is observed at the start of cycle 23 is that
the rush started at all altitudes early in 1997 when it branched away
from the previous equatorward drift, and intersected the start of the
next equatorward drift at the start of 2000. As with the equatorward
drift, we have indicated our estimates of the rush to the poles during
the rising phase of cycle 23 with dotted black and white lines in
Figure~\ref{fig:activity96}. The start of the rush to the poles of
cycle 24 is also apparent in the green line and the LASCO C2 data for
2009 and 2010.

\section{Discussion}
\label{sec:discussion}

The most obvious trend that we see as we track the zones of activity to
higher altitudes is that the drift towards the equator becomes more
rapid as we move higher into the corona. 

Could this be an observational artifact caused by changes in the
Thomson-scattering weighting functions?  For this to be the case, it
would be necessary that the Thomson weighting functions become narrower
(in angle as seen from the Sun) as the closest approach of the line of
sight to the Sun moves further from the Sun. In fact the reverse is the
case as the electron density falls off faster than $1/r^2$ close to the
Sun (\textit{e.g.} \citealp{allen73}). In addition, the MLSO data (heights
below 2.35\rsun{}) use polarised light which has a much narrower
weighting function than unpolarised (\textit{e.g.}\ \citealp{howard09}). The
emission-line weighting is generally narrower than the Thomson function
at the same distance since the dependence of the emission is
approximately proportional to $N_\mathrm{e}^{1.7}$ and peaks at a temperature of
about $1.8\times10^6$K (\citealp{mason75}; \citealp{guhathakurta92}).
We have verified this by some simple simulations of radial
streamers. These show that such streamers can never appear at a lower
projected latitude than their footpoint, and that except at very high
footpoint latitudes (above about 60\degr{}) they predominantly appear
close to the footpoint latitude.
If (as
seems probable) we are looking at a band of activity reaching up to a
particular latitude, then a wider scattering function will increase the
sensitivity of the observations to structures well-away from the sky
plane. Since a radial streamer away from the sky plane will have an
apparent latitude greater than its true latitude this would tend to
make that activity appear to extend closer to the poles, which is the
reverse of the trend that we see, thus confirming that it cannot be an
artifact of the different observing techniques.

It is well-known that flows from coronal holes diverge (\textit{e.g.}\
\citealp{whang83}; \citealp{falceta05}), and also that during the
declining and minimum phase of the solar cycle the solar wind is
well-represented as a two-part system with low-latitude slow wind from
the streamer belt and high-latitude fast wind coming from the polar
coronal holes (\textit{e.g.}\ \citealp{phillips95}).  The behaviour of the
equatorward drift may then be understood if we assume that the upper
zone of activity lies on the boundary between the low latitude corona
and the polar coronal holes (most probably streamers overlying the
polar crown filaments). That boundary moves equatorward as the
polar holes become larger during the decline of the solar cycle. The
divergence of the flow from the holes also squeezes the streamers
towards the equator at higher altitudes. When the polar hole is smaller
and weaker near solar maximum, we expect that not only will the
boundary be closer to the poles, but that there will also be less
divergence. While the exact degree of divergence will depend on the
speeds and densities of the flows in the coronal holes and the streamer
belt, the degree of equatorward deflection which we see is consistent
with that found the in the MHD simulations of \citet{whang83}.

\begin{figure}[tbp]
  \centering
  \includegraphics[width=\textwidth]{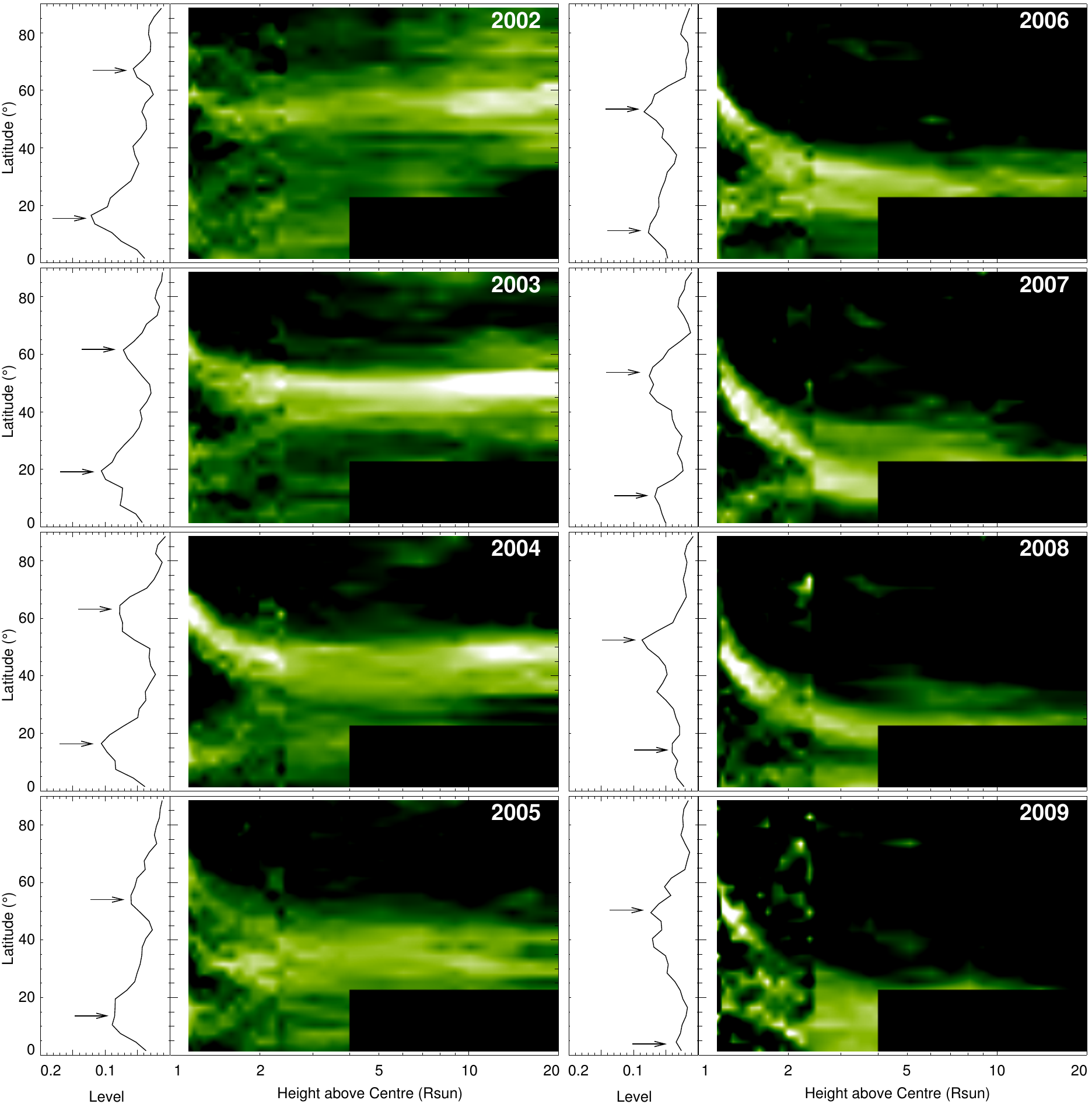}
  \caption{Annual averages of the activity distribution in the K-corona
    for the declining phase of cycle 23. The green line activity
    profile is plotted to the left of each year. The approximate
    latitudes of the two bands of activity are marked by the arrows on
    the green line profiles. LASCO C3 data below 20\degr{} latitude are
    omitted as these reflect the F-corona rather than the K-corona. For
    2009, the low altitude data from MLSO are noisy as there were many
    data gaps during that year.}
  \label{fig:profiles_1}
\end{figure}

To better visualise the shape of the activity zones in latitude and
radius, we made annual averages of the activity levels as a function of
radius and latitude for the declining phase of cycle 23, these are
shown in Figure~\ref{fig:profiles_1}.  In this format the image is a
representation of the average path of the streamers from close to the
Sun, into the high corona.  The deflection of the activity zones
towards the equator with increasing height in the low corona is very
evident, as is their more radial nature in the higher corona. It is also
very clear that the amount of deflection increases as we approach solar
minimum.

We also see that the feature corresponding to the low-latitude
``current'' cycle activity zone in the low corona cannot be traced
above about 3\rsun{} after 2003. It is unclear whether the streamers
overlying the main activity belt simply do not extend into the high
corona or whether they deflect polewards and merge with the
high-latitude zone. The latter is hinted by some years, notably 2006
and 2007, however the apparent band joining the low-latitude activity
to the high-latitude is in the MLSO data which have significantly
poorer quality than either the NSO green line data or the LASCO data.

\begin{figure}[tbp]
  \centering
  \includegraphics[width=\textwidth]{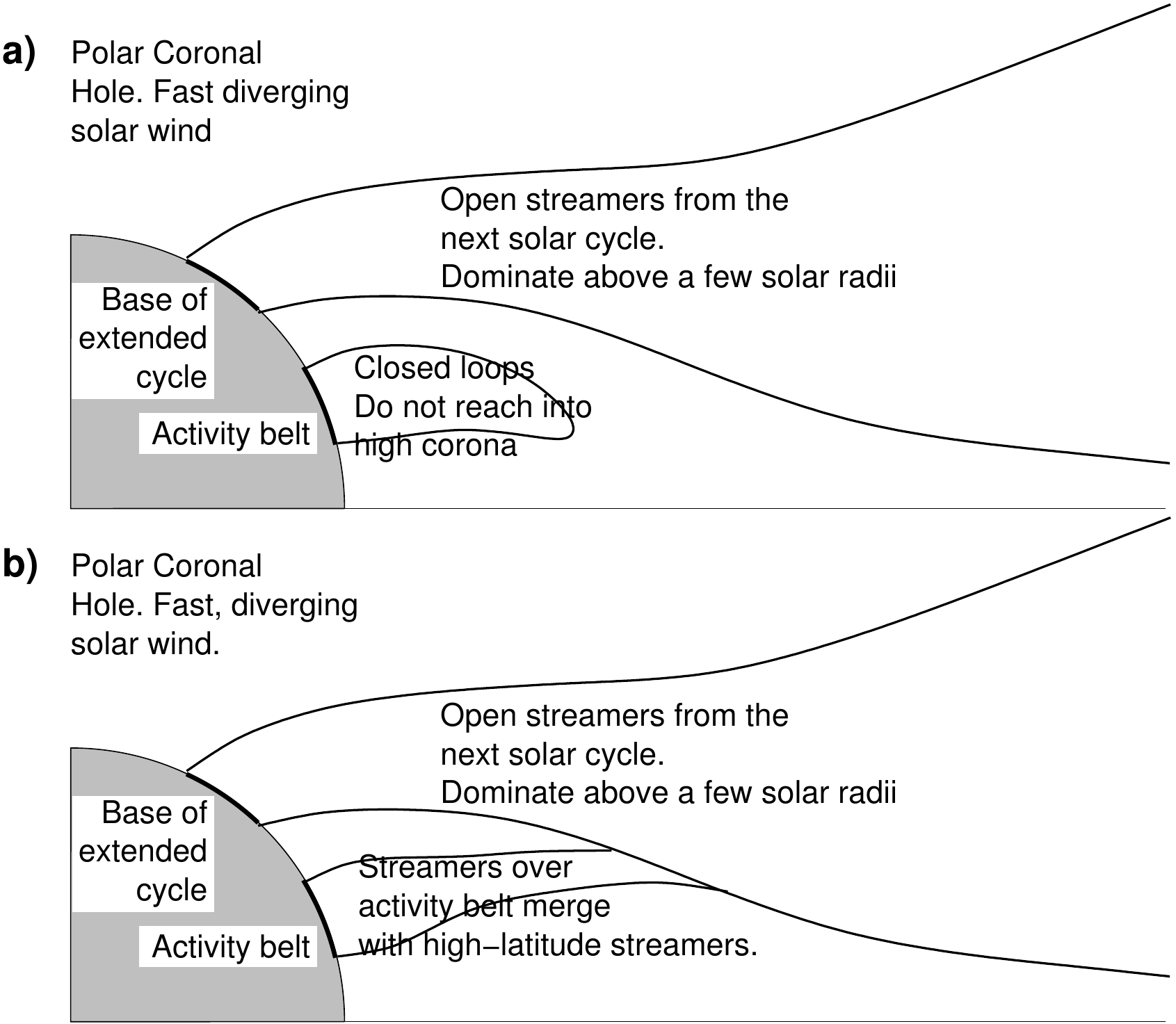}
  \caption{Schematic illustration of two possible interpretations of
    the streamer structure in the late declining phase of the solar
    cycle. (a) The streamers over the main activity belt are closed
    loops and do not extend above a few \rsun{}, (b) The activity belt
    streamers merge with the high-latitude streamers.}
  \label{fig:streamer-schematic}
\end{figure}

We thus have two possible interpretations of the long-term structure of
the coronal activity during the declining phase of the cycle. These are
illustrated schematically in Figure~\ref{fig:streamer-schematic}. If
the apparent connection of the two bands at about 2\rsun{} is an
artifact, then the structure is as shown in
Figure~\ref{fig:streamer-schematic}(a), where the streamers over the
main activity zone are closed loops which do not extend above a few
solar radii. If on the other hand the connection is real, then the
structure must be as in Figure~\ref{fig:streamer-schematic}(b), and the
streamers from the activity belt merge with those originating at higher
latitudes. It is of interest to compare this with the interpretation of
a coronal ray observed during a solar minimum eclipse shown in
Figure~136 of \citet{shklovskii65}.

Irrespective of which interpretation of the topology is correct, it is
evident that at higher altitudes ($>3\rsun$) the streamer belt is
predominantly rooted in the higher-latitude activity band closer to the
Sun. Hence we must expect that the slow solar wind seen by
\textit{in-situ} observations near Earth during the declining phase of
the solar cycle, and especially after the lower edge of the equatorward
drift reaches the equator (around 2007 for cycle 23), will include
material characteristic of the next solar cycle.

We conjecture that the high-latitude streamers are those originating
over the polar crown filaments, while the low-latitude ones originate
above active regions. If this is the case then it is to be expected
that as activity declines to solar minimum the importance of streamers
from the active regions will decrease relative to those from the polar
crown. However a full understanding of this would require a detailed
study of the rising phase of the solar cycle, including photospheric
and chromospheric observations to show how the polar-crown region
transitions to the activity-belt region.

As with the equatorward drift, we find that the highest latitudes of the
rush to the poles is similar at all heights. However at the start of
the rush, it is at much lower latitudes at higher altitudes. It
branches away from the equatorward drift at similar times at all
altitudes. This may be explained in the same way as the latitudinal
variation of the equatorward drift with altitude.  That is, the polar
coronal hole shrinks during the rise in activity, and so the boundary
between the polar hole region and the polar crown region moves
poleward and the flows from the polar hole become weaker. Thus as the
rush move poleward its equatorward deflection decreases.

It is also evident from Figure~\ref{fig:trends}a that for those heights
at which we were able to determine the rush to the poles for cycle 24,
it was taking place more slowly than at the corresponding heights in
cycle 23. However the uncertainties in the determination of the rush
for cycle 24 are large, and it is not clear that this is actually
saying anything other than that the time between the maximum of cycle 23
and that of cycle 24 is longer than the interval in cycles 22 and 23.

\section{Conclusions}
\label{sec:conclusions}

We have shown that the activity zones of the extended solar cycle seen
in the emission corona at 1.15\rsun{} can be traced far out into the
corona in the Thomson-scattered light of the K-corona. We find that the
activity zones are deflected towards the equator at greater heights in
the corona. We also see that as the activity moves toward the equator
through the declining phase of the cycle, the amount of deflection
increases. This is consistent with the expansion of the flows from the
polar coronal holes. This deflection also seen in the rush to the poles
during the rising phase of the cycle.  

During the declining phase of
the solar cycle, the activity zones above a few solar radii are
dominated by structure which connect to the high-latitude branch in the
green-line activity, which is related to the upcoming solar cycle.
It therefore seems inevitable that at least in the late declining phase
of the solar cycle the low-latitude slow solar wind is more
characteristic of the upcoming cycle than of the current. This opens up
the possibility of determining some characteristics of the next cycle
from \textit{in-situ} measurements made well before the start of that
cycle. In the case of the current cycle, it appears that cycle 24
should dominate from about 2007.

\subsubsection*{acknowledgements}
  The Mauna Loa coronameter data are provided courtesy of the Mauna Loa
  Solar Observatory, operated by the High Altitude Observatory, as part
  of the National Center for Atmospheric Research (NCAR). NCAR is
  supported by the National Science Foundation. LASCO was built by a
  consortium of the Naval Research Laboratory (Washington, USA), the
  Max-Planck-Institut f\"ur Aeronomie (Lindau, Germany) the Laboratoire
  d'Astronomie Spatiale (Marseille, France) and the University of
  Birmingham (Birmingham, UK) and is a part of the SOHO mission which
  is a collaborative mission of ESA and NASA. Partial support for NSO
  (including the work of SJT and the operation of the coronagraph) is
  provided by the US Air Force under a Memorandum of Agreement.

\end{document}